\documentclass[a4paper,11pt]{article}

\usepackage{jcappub} 

\usepackage[T1]{fontenc} 

\usepackage{amssymb}
\usepackage{epsfig} 
\usepackage{amsmath} 
\usepackage{graphicx}

\newcommand{\beq}{\begin{equation}}
\newcommand{\eeq}{\end{equation}}
\newcommand{\bea}{\begin{eqnarray}}
\newcommand{\eea}{\end{eqnarray}}
\newcommand{\ba}{\begin{array}}
\newcommand{\ea}{\end{array}}
\newcommand{\bi}{\begin{itemize}}
\newcommand{\ei}{\end{itemize}}
\newcommand{\bn}{\begin{enumerate}}
\newcommand{\en}{\end{enumerate}}
\newcommand{\bc}{\begin{center}}
\newcommand{\ec}{\end{center}}
\renewcommand{\l}{\left}
\renewcommand{\r}{\right}

\newcommand{\eq}[1]{Eq.~(\ref{#1})}
\newcommand{\eqs}[2]{Eqs.~(\ref{#1}) and (\ref{#2})}

\newcommand{\eV}{\mathinner{\mathrm{eV}}}

\newcommand{\MeV}{\mathinner{\mathrm{MeV}}}
\newcommand{\GeV}{\mathinner{\mathrm{GeV}}}
\newcommand{\TeV}{\mathinner{\mathrm{TeV}}}

\newcommand{\pl}{{\rm Pl}}

\newcommand{\axino}{{\tilde{a}}}
\newcommand{\sneu}{{\tilde{N}}}

\title{\boldmath PQ-symmetry for a small Dirac neutrino mass, dark radiation and cosmic neutrinos}

\author{Wan-Il Park}

\affiliation{School of Physics, KIAS, \\ Seoul 130-722, Korea}

\emailAdd{wipark@kias.re.kr}

\abstract{
We propose a supersymmetric scenario in which the small Yukawa couplings for the Dirac neutrino mass term are generated by the spontaneous-breaking of Pecci-Quinn symmetry.
In this scenario, a right amount of dark matter relic density can be obtained by either right-handed sneutrino or axino LSP, and a sizable amount of axion dark radiation can be obtained.
Interestingly, the decay of right-handed sneutrino NLSP to axino LSP is delayed to around the present epoch, and can leave an observable cosmological background of neutrinos at the energy scale of $\mathcal{O}(10-100) \GeV$.
}

\begin{document}
\maketitle
\flushbottom

\section{Introduction}

The standard model (SM) has been extremely successful in describing subatomic world, but many astrophysical and cosmological observations require a theory beyond SM.
One of the apparent shortcomings of SM is the lack of the tiny neutrino mass hinted by atmospheric and solar neutrino oscillations \cite{Fukuda:1998mi,Ahmad:2001an} (also reactor and long-baseline neutrino oscillations \cite{Eguchi:2002dm,Michael:2006rx}).
The mass-squared differences of neutrino mass eigenstates are now known to be \cite{Fogli:2012ua}
\beq
|\Delta m_{21}^2| \simeq 7.5 \times 10^{-5} \eV^2, \quad |\Delta m_{32}^2| \simeq 2.4 \times 10^{-3} \eV^2
\eeq
This implies that at least one neutrino has a mass of at least $0.05 \eV$.
On the other hand, a recent analysis based on data from Planck satellite mission \cite{planck} and predictions from other phenomena found a consistent picture of $\Lambda$CDM model with the sum of the active neutrino masses given by \cite{Battye:2014}
\beq
\sum m_\nu = 0.320 \pm 0.081 \eV
\eeq
When one tries to get such small neutrino masses from the Higgs mechanism with right-handed (RH) neutrinos introduced, very tiny Yukawa couplings of $\mathcal{O}(10^{-13}-10^{-12})$ are required. 
It looks quite puzzling to have such small Yukawa couplings.
The best known mechanism for this puzzle is the so-called seesaw mechanism \cite{GellMann:1980vs,Yanagida:1979as,Mohapatra:1979ia} which uses a large Majorana mass term of RH-neutrinos.
However one should note that a Majorana particle has never been observed so far and the small Yukawa couplings may have a dynamical origin.

Besides phenomenological issues, SM also suffers from an esthetic theoretical issue, strong CP problem \cite{Kim:1986ax,Peccei:2006as} requiring a tuning of $\mathcal{O}(10^{-10})$ to match experimental data \cite{Dress:1976bq,Altarev:1996xs}.
Axion \cite{Peccei:1977hh,Peccei:1977ur,Weinberg:1977ma,Wilczek:1977pj,Kim:1979if,Shifman:1979if,Zhitnitsky:1980tq,Dine:1981rt} from the breaking of a global Abelian symmetry (called $U(1)_{\rm PQ}$) \cite{Peccei:1977hh,Peccei:1977ur} provides a very simple and attractive solution to this problem, while becoming a good candidate of cold dark matter.
Additionally, as we discuss in this paper, the symmetry breaking field associated with $U(1)_{\rm PQ}$ may be responsible for the small Dirac neutrino mass term.

Meanwhile, low energy supersymmetry (SUSY) is quite attractive because it can provide a fine unification of SM gauge couplings, a natural solution to the hierarchy problem of electroweak Higgs mass and a candidate of dark matter under the assumption of $R$-parity conservation, even though it is facing with \textit{little hierarchy problem} arising due to the lack of SUSY signature at recent collider experiments \cite{Feng:2013pwa}.
In particular, dark matter might be from the extended non-MSSM sector which is necessary to address various theoretical/phenomenological shortcomings of SM.

In this paper, we propose a supersymmetric extension of SM in which the tiny Yukawa couplings of the Dirac neutrino mass term are generated dynamically by Peccei-Quinn field which breaks $U(1)_{\rm PQ}$ symmetry spontaneously, and discuss its cosmological implications including dark matter, dark radiation and cosmic neutrino flux.

This paper is organized as follows.
In section~\ref{sec:model}, our model is described.
In section~\ref{sec:cosmo}, cosmological aspects of scalar fields (particularly, RH-sneutrino and saxion fields) are briefly discussed.
In section~\ref{sec:DM}, we estimate abundances of LSP and NLSP before it decays.
In section~\ref{sec:DR}, the dark radiation contribution of relativistic axions produced in the decay of saxion is estimated.
In section~\ref{sec:neu-flux}, a cosmological neutrino flux produced in the decay of NLSP is estimated.
In section~\ref{sec:con}, our conclusion is provided.

\section{The model} \label{sec:model}
We consider the following superpotential,
\beq \label{W}
W = W_{{\rm MSSM}-\mu} + \lambda_\mu \frac{X^2}{M_*} H_u H_d + \lambda_\nu \l( \frac{X}{M_*} \r)^2 L H_u N^c + \lambda_\Psi X \Psi \bar{\Psi} 
\eeq
where gauge structure and family indices were suppressed, 
\beq
W_{{\rm MSSM}-\mu} = y_u Q H_u u^c + y_d Q H_d d^c + y_\ell L H_d e^c
\eeq
is the SUSY-extension of SM Yukawa interactions, $M_*$ is a UV-cutoff scale where associated higher order operators become effective, $X$ is a SM-gauge singlet, $N$ is the right-handed neutrino super field, and $\Psi$ ($\bar{\Psi}$) is assumed to be a complete representation of $SU(5)$ to maintain gauge-unification.
 The underlying symmetries for the superpotential \eq{W} are assumed to be $U(1)_{\rm PQ}$ and $U(1)_{L}$ under which charges are assigned as Table~\ref{tab:charges} 
 \footnote{
Alternatively, one can use only $U(1)_{\rm PQ}$ symmetry with charges assigned as $q_{\rm PQ}(LH_u,N)=(-5/2,-1/2)$ with all the others same as Table~\ref{tab:charges}.
In this case, a higher order term like $\lambda_N (N^c)^4/M_*$ is not allowed since PQ-symmetry is supposed to be accurate up to a very high order in order not to spoil the axion solution to the strong CP problem \cite{Kamionkowski:1992mf}.}.
\begin{table}[htdp]
\begin{center}
\begin{tabular}{|c|c|c|c|c|c|c|c|}
\hline
 & $X$ & $H_uH_d$ & $LH_u$ & $N$ & $\Psi \bar{\Psi}$ \\
 \hline
 $q_{\rm PQ}$ & $1$ & $-2$ & $-2$ & $0$ & $-1$ \\
 \hline
 $q_{L}$ & $0$ & $0$ & $1$ & $1$ & $0$ \\
 \hline
\end{tabular}
\end{center}
\caption{\label{tab:charges} Charge assignment for $U(1)_{\rm PQ}$ and $U(1)_L$ symmetries.}
\end{table}

When $X$ develops a vacuum expectation value (VEV), $X_0$, MSSM $\mu$-term can be reproduced as
\beq
\mu = \lambda_\mu \frac{X_0^2}{M_*}
\eeq
Simultaneouly, the Yukawa coupling of the Dirac neutrino mass term can be generated, leading to the Dirac neutrino mass given by
\beq
m_\nu^{\rm D} = \lambda_\nu \l( \frac{X_0}{M_*} \r)^2 v \sin \beta
\eeq
where $v \equiv \sqrt{v_u^2 + v_d^2} = 174 \GeV$ with $v_{u,d}$ being the VEV of the neutral component of $H_{u,d}$ and $\tan \beta = v_u/v_d$.
Data from collider experiments may imply $\mu = \mathcal{O}(10^{2-3}) \GeV$ and observations requires $\sum m_\nu^{\rm D} \lesssim 0.3 \eV$ \cite{Battye:2014} which can be translated to 
\beq
X_0 \simeq \frac{7.6 \times 10^{-7} M_*}{\sqrt{\lambda_\nu \sin \beta}} \l( \frac{m_\nu^{\rm D}}{0.1 \eV} \r)^{1/2} 
\eeq
where we assumed $\lambda_\nu$ to be diagonal for simplicity.
We may take $M_* = M_{\rm GUT} \simeq 2 \times 10^{16} \GeV$, and assume  
\beq
\lambda_\mu \ , \lambda_\nu \sim \mathcal{O}(10^{-3} -1)
\eeq
which leads to 
\beq
X_0 = \mathcal{O}(10^{10-11}) \GeV
\eeq
Note that $X$ can be stabilized at such an intermediate scale by the radiative running of soft mass-squared of $X$ (thanks to $\lambda_\Psi$ term in \eq{W}) or by the interplay between the negative soft mass-squared term and a dimension-six term (e.g., $|X|^6$ term) in the scalar potential of $X$ if we include a term, for example, $X^3 Y/M_*$ in the superpotential
\footnote{For the term $X^3 Y/M_*$, $Y$ should be another SM-gauge-singlet with a PQ-charge, $-3$ for axion solution to the strong CP-problem.}.
This is how a tiny Dirac neutrino mass is obtained in a natural manner without resorting to an extremely small $\lambda_\nu$ or large Majorana mass.

The particle spectra in the PQ-sector involving the symmetry breaking field $X$ depends on how $X$ is stabilized.
For example, if $X$ is stabilized radiatively, the mass of saxion (the radial component of $X$) is smaller than soft SUSY-breaking mass of $X$ by an order of magnitude, and axino (the fermonic super-partner of $X$) can be quite light, having a mass of $\mathcal{O}(1-10) \GeV$ \cite{Kim:2008yu}.
On the other hand, if the stabilization is achieved by a higher order term, all the particles (except axion) in the PQ-sector have masses similar to the soft mass of $X$.
Here, we do not specify stabilization mechanism, but keep in mind those two possibilities;
\beq
m_\axino \ll m_{\rm PQ} \ll m_X \lesssim m_{\rm soft}, \ {\rm or} \ m_\axino \sim m_{\rm PQ} \sim m_X \lesssim m_{\rm soft}
\eeq
where $m_\axino, m_{\rm PQ}, m_X$ and $m_{\rm soft}$ are respectively the mass of axino, saxion, soft SUSY-breaking mass of $X$, and the typical scale of soft SUSY-breaking mass.

\section{Cosmology} \label{sec:cosmo}
In this section, we discuss briefly cosmological evolutions of RH-sneutrino and saxion.
For simplicity, we assume Planck-scale moduli are heavy enough and do not have any significant effects on our argument.
    
\subsection{RH-sneutrino coherent osciallation}
RH-sneutrino field ($\sneu$) might be held around the origin during and after inflation.
However, it is also possible to have a non-zero VEV of $\sneu$ during inflation.
For example, there can be interactions of inflaton to other field(s) in K\"{a}hler potential. 
In the presence of a coupling to inflaton, $\tilde{N}$ can obtain a mass term of Hubble scale tachyonic mass \cite{Dine:1995uk}.
Since gravitational effects can break a global symmetry explicitly, we may add a term like $\lambda_N (N^c)^4/(4 M_*)$ in \eq{W}, and find out the field value $\sneu_0 \sim \l( \frac{m_{\tilde{N}} M_*}{\sqrt{3} \lambda_N} \r)^{1/2}$ at the onset of the coherent oscillation taking place as $H \lesssim m_\sneu$.
It may also be possible that lepton number is quite good symmetry, pushing up the symmetry breaking operator to a very high order. 
In this circumstance, the initial oscillation amplitude of RH-sneutrino, $\sneu_0$, can be treated as a free parameter.
Then, the late-time abundance of RH-sneutrino which is associated with the coherent oscillation can be expressed as
\beq \label{Y-RHsneu-ini}
Y_{\tilde{N}} \sim \l( \frac{\tilde{N}_0}{M_\pl} \r)^2 \frac{T_{\rm R}}{m_{\tilde{N}}}
\eeq
where $T_{\rm R}$ is the reheating temperature after primordial inflation, and \eq{Y-RHsneu-ini} is valid for $\mathcal{O}(10) \MeV \lesssim T_{\rm R} \lesssim \sqrt{m_{\tilde{N}} M_\pl}$.

\subsection{Thermal inflation}
Similarly to the case of RH-sneutrino field, flaton field $X$ may or may not be held around the origin.  
If $X$ were destabilized during inflation due to negative Hubble mass term, the coherent oscillation of $X$ starts as $H \lesssim m_X$ with an amplitude of order of $X_0$.
If $X$ were held around the origin due to the interaction with thermal bath via the $\lambda_\Psi$ term, there is a chance to have a phase of thermal inflation \cite{Lyth:1995hj,Lyth:1995ka} having about $10$ $e$-foldings.
Thermal inflation eventually ends as $X$ is destabilized at $T_{\rm c} \sim m_X$, and the following coherent oscillation of $X$ dominates the energy density of the universe.

The condensation of $X$ eventually decays to SM particles, thanks to the $\mu$-term interaction.
The temperature when flatons decay can be defined as
\beq
T_{\rm d} \equiv \l( \frac{\pi^2}{90} g_{*S}(T_{\rm d}) \r)^{-1/4} \sqrt{\Gamma_X M_\pl}
\eeq
where $\Gamma_X$ is the total decay rate of $X$.
One finds 
\beq \label{GammaX}
\Gamma_X \simeq \Gamma_{X \to aa} + \Gamma_{X \to \rm SM}
\eeq
where \cite{Kim:2008yu}
\bea \label{Xtoaa}
\Gamma_{X \to aa} &=& \frac{1}{64 \pi} \frac{m_{\rm PQ}^2}{X_0^2}
\\ \label{XtoSM}
\Gamma_{X \to \rm SM} &\sim& 16 \l( 1 - \frac{|B|^2}{m_A^2} \r)^2 \l( \frac{\mu}{m_{\rm PQ}} \r)^4 f(m_h^2/m_{\rm PQ}^2) \Gamma_{X \to aa}
\eea
with
\beq
f(x) = \sqrt{1-4 x} + \frac{\epsilon x}{(1-x)^2} + \l( 1- \frac{\epsilon x}{3} \r)^{3/2}
\eeq
and $\epsilon \sim 12 m_b^2/m_h^2$.
In case of thermal inflation, the decay of flaton is the main source of the radiation background having temperature $T_{\rm d}$, and releases a huge amount of entropy.
As a result, the abundance of particles pre-existing before thermal inflation is diluted by a factor
\beq \label{TI-dilution}
\Delta \equiv \frac{s_{\rm after}}{s_{\rm before}} \simeq \frac{30}{\pi^2 \ g_{*S}(T_{\rm c})} \frac{V_0}{T_{\rm c}^3 T_{\rm d}}
\eeq
where $s_{\rm before}$ and $s_{\rm after}$ are the entropy densities before and after the completion of flaton decay in the sudden decay approximation, and $V_0 \sim m_{\rm PQ}^2 X_0^2$ is the energy density of thermal inflation.

The very large dilution factor $\Delta$ in thermal inflation is problematic for most baryogenesis scenarios and requires a particular mechanism \cite{Stewart:1996ai,Jeong:2004hy,Kawasaki:2006py,Felder:2007iz,Choi:2009qd,Park:2010qd}.
However, thermal inflation may or may not exist and it is out of scope of this paper to discuss the possibility of baryogenesis.
So, here we simply assume that there is a working baryogenesis mechanism.

\section{Dark matter relic density} \label{sec:DM}
We assume an axino LSP and RH-sneutrino NLSP scenario in this paper.
%
In this case, axinos can be produced mainly from decays of flaton and neutralino in thermal bath, as well studied in Ref.~\cite{Kim:2008yu}.
RH-sneutrino can be produced from the decay of neutralino too, but flaton contribution is negligible due to very small coupling.
As described in section~\ref{sec:cosmo}, RH-sneutrino also has a coherent production mechanism which can be its main production channel.
For later use, we collect the decay rates of neutralino and NLSP as follows.
\bea \label{chi-to-axino}
\Gamma_{\chi \to \tilde{a} + \rm SM} 
&=& \frac{\gamma_\chi}{16 \pi} \frac{m_\chi^3}{X_0^2}
\nonumber \\
&\simeq& 2 \times 10^{-15} \GeV \gamma_\chi \l( \frac{m_\chi}{1 \TeV} \r)^2 \l( \frac{10^{11} \GeV}{X_0} \r)^2
\\ \label{chi-to-sneutrino}
\Gamma_{\chi \to \tilde{N} + \nu} 
&\simeq& \frac{1}{16 \pi} |\lambda_{\nu, \rm eff} \Theta_{\tilde{H}_u}^\chi|^2 \l( 1 - \frac{m_\sneu^2}{m_\chi^2} \r)^2 m_\chi
\nonumber \\
&\simeq& 8.0 \times 10^{-25} \GeV \l( \frac{\lambda_{\nu, \rm eff}}{2 \times 10^{-13}} \r)^2 \l( \Theta_{\tilde{H}_u}^\chi \r)^2 \l( 1 - \frac{m_{\tilde{N}}^2}{m_\chi^2} \r)^2 \l( \frac{m_\chi}{1 \TeV} \r) 
\\ \label{sneutrino-to-axino}
\Gamma_{\sneu \to \axino + \nu} 
&\simeq& \frac{1}{2 \pi} |\lambda_{\nu, \rm eff}|^2 \l( \frac{v \sin \beta}{X_0} \r)^2 \l( 1 - \frac{m_{\tilde{a}}^2}{m_{\tilde{N}}^2} \r)^2 m_\sneu
\nonumber \\ 
&\simeq& 2 \sin^2 \beta \times 10^{-41} \GeV \l( \frac{10^{11} \GeV}{X_0} \r)^2 \l( 1 - \frac{m_\axino^2}{m_\sneu^2} \r)^2 \l( \frac{m_\sneu}{1 \TeV} \r) 
\eea 
where $\gamma_\chi \sim \mathcal{O}(1)$ including various channel-dependence and the phase space factor, $\lambda_{\nu, \rm eff} \equiv \lambda_\nu \l( X_0 / M_* \r)^2$, $\Theta_{\tilde{H}_u}^\chi$ is the fraction of the lightest neutralino in the Higgsino $\tilde{H}_u$, and the mass of neutrino was ignored.
As can be seen from \eq{sneutrino-to-axino}, the life time of the NLSP can be close to the age of Universe.
This means that the abundance of NLSP can be constrained by cosmic background of neutrinos \cite{Aartsen:2012uu}.

In the following discussion, we use the term ``\textit{symmetric phase}'' for the case in which flaton field ($X$) is held around origin in the very early universe and thermal inflation takes place, and ``\textit{broken phase}'' for the case in which $X$ is destabilized during inflation and thermal inflation does not take place.

\subsection{Axinos}
Axino production in the symmetric phase was well described in Ref.~\cite{Kim:2008yu}.
So, here we borrow only the results there, and add the case of broken phase (see also Ref.~\cite{Chun:2011zd,Bae:2011jb,Choi:2011yf,Bae:2011iw}).

\subsubsection{Axinos from the flaton decay}
If flaton is stabilized by an higher order operator, the mass of axino is comparable to that of saxion, and the decay of saxion to axinos are kinematically forbidden.
On the other hand, if the stabilization is achieved by the radiative running of the soft SUSY-breaking mass-squared parameter, axino can be much lighter than saxion, and the decay rate of flaton to axino can be expressed as \cite{Kim:2008yu}
\beq \label{Xtoaxino}
\Gamma_{X \to \axino \axino} = \frac{\alpha_\axino^2 m_\axino^2 m_{\rm PQ}}{32 \pi X_0^2}
\eeq
where the mass of axino was ignored in the phase space factor.
Then, in the symmetric phase, the fractional energy density of axinos at the present universe is given by \cite{Kim:2008yu}
\beq \label{flaton-axino}
\Omega_\axino 
\simeq 0.36 \frac{\Gamma_X^{1/2}}{\Gamma_{X \to \rm SM}^{1/2}} \l( \frac{10}{g_*^{1/2}(T_{\rm d})} \r) \l( \frac{\alpha_\axino}{0.1} \r)^2 \l( \frac{m_\axino}{1 \GeV} \r)^3 \l( \frac{10 \GeV}{T_{\rm d}} \r) \l( \frac{10^{11} \GeV}{X_0} \r)^2 
\eeq

In the broken phase, from the axino number density given by
\beq
n_\axino = \frac{2 \Gamma_{X \to \axino \axino}}{m_{\rm PQ} a(t)^3} \int_0^t a(t')^3 \rho_X(t') dt' 
\eeq
the late time abundance of axino is obtained as
\beq
Y_\axino = \frac{3}{2} e^{\Gamma_X t_{\rm d}}  \frac{T_{\rm d}}{m_{\rm PQ}} \frac{\Gamma_{X \to \axino \axino}}{\Gamma_X} \frac{\rho_X(t_{\rm d})}{\rho_{\rm r}(t_{\rm d})}
\eeq
where $\rho_X(t_{\rm d})$ and $\rho_{\rm r}(t_{\rm d})$ are respectively the energy densities of flaton and background radiation at $t=t_{\rm d} \equiv 1/\Gamma_X$.
Hence, from Eqs.~(\ref{GammaX}),~(\ref{Xtoaa}),~(\ref{XtoSM}),~(\ref{Xtoaxino}) and 
\bea
\label{dilution-I}
\frac{\rho_X(t_{\rm d})}{\rho_{\rm r}(t_{\rm d})} &\sim& \l( \frac{m_{\rm PQ}}{m_X} \r)^2 \l( \frac{X_0}{M_\pl} \r)^2 \frac{T_{\rm R}}{T_{\rm d}}
\eea
where $\mathcal{O}(10) \MeV \lesssim T_{\rm R} \lesssim \sqrt{m_X M_\pl}$, the fractional energy density of axino at present is given by 
\bea \label{no-TI-flaton-axino}
\Omega_\axino 
&=&  3.2 \times 10^{-13} \l( \frac{\alpha_\axino}{0.1} \r)^2 \l( \frac{m_\axino}{1 \GeV} \r)^3 \l( \frac{500 \GeV}{m_X} \r)^2 \l( \frac{T_{\rm R}}{m_{\rm PQ}} \r) \l( \frac{X_0}{10^{11} \GeV} \r)^2
\nonumber \\
&& \times \l[ 1 + 16 \l( 1 - \frac{|B|^2}{m_A^2} \r)^2 \l( \frac{\mu}{m_{\rm PQ}} \r)^4 f(m_h^2/m_{\rm PQ}^2) \r]^{-1}
\eea

\subsubsection{Axinos from the neutralino decay}
In the symmetric phase, when 
\beq
\frac{2}{3} \frac{g_*(T_{\rm d})^{1/4}}{g_*(T_\chi)^{1/4}} \frac{T_{\rm d}}{T_\chi} \ll 1
\eeq
with $T_\chi \equiv (2/21) m_\chi$, using \eq{chi-to-axino}, one finds that the energy contribution of axinos from decays of neutralinos is given by \cite{Kim:2008yu}
\beq \label{neutralino-axino}
\Omega_{\tilde{a}} \sim 0.19 \gamma_\chi \frac{\Gamma_{\rm SM}^{1/2}}{\Gamma_X^{1/2}} \l( \frac{10^3 g_*(T_{\rm d})^{3/2}}{g_*(T_\chi)^3} \r) \l( \frac{m_\chi}{100 \GeV} \r) \l( \frac{m_{\tilde{a}}}{1 \GeV} \r) \l( \frac{10^{11} \GeV}{X_0} \r)^2 \l( \frac{T_{\rm d}}{m_\chi/25} \r)^7
\eeq

In the broken phase, one can use the Boltzmann equation leading to 
\beq
Y_\axino \simeq \frac{135 g_\chi}{4 \pi^4 g_{*S}(T_{\rm R}) g_*^{1/2}(T_{\rm R})} \frac{M_\pl \Gamma_{\chi \to \axino + \rm SM}}{m_\chi^2} \int_{x_{\rm R}}^\infty x^3 K_1(x) dx
\eeq
where $x \equiv m_\chi/T$ and $K_1(x)$ is the first modified Bessel Function of the 2nd kind.
For $T_{\rm fz} \lesssim T_{\rm R} \ll m_\chi$ (or $1 \ll x_{\rm R} \lesssim x_{\rm fz}$), the integral is approximated to $\sqrt{\pi/2} x_{\rm R}^{5/2} \exp(-x_{\rm R})$.
Hence the fractional energy density of axinos at present is obtained to be
\beq \label{neutralino-axino-no-TI}
\Omega_\axino \simeq 0.265 \gamma_\chi \l( \frac{m_\axino}{1 \GeV} \r) \l( \frac{m_\chi}{500 \GeV} \r) \l( \frac{10^{11} \GeV}{X_0} \r)^2 \l( \frac{x_{\rm R}}{15} \r)^{5/2} {\rm Exp}\l[-x_{\rm R} + 15 \r]
\eeq
where $g_\chi=2$ and $g_{*S}(T_{\rm R}) = g_*(T_{\rm R}) = 100$ were used.
This contribution is dominant over the one from flaton decay (\eq{no-TI-flaton-axino}) for which axino mass is likely to be smaller than flaton mass by about an order of magnitude.

\begin{figure}[h]
\centering
\includegraphics[width=0.45\textwidth]{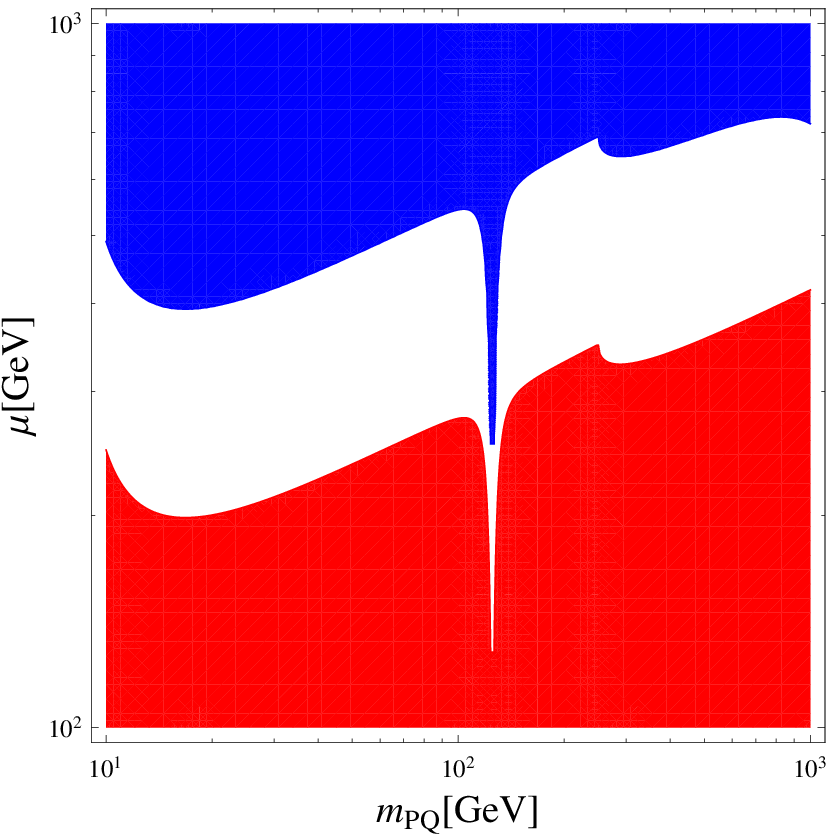}
\includegraphics[width=0.45\textwidth]{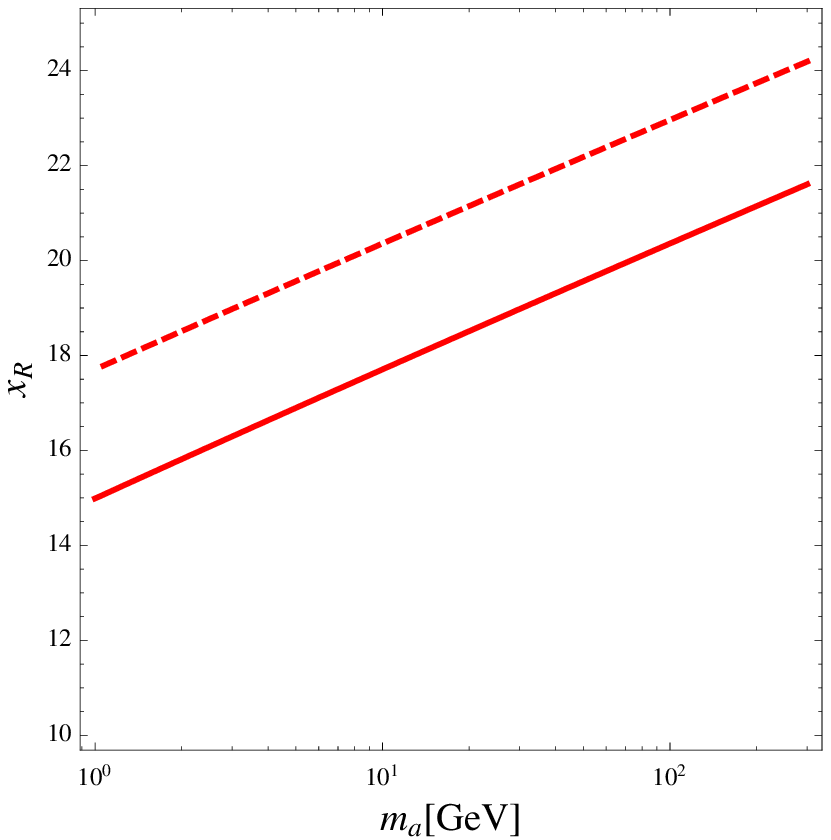}
\caption{\label{fig:axinos} 
Relic abundance of axinos.
Left: Symmetric phase. $\gamma_\chi=1$, $|B|=0.8 m_A$, $m_\axino = 1 \GeV$, $m_\chi = 300 \GeV$, $X_0 = 2 \times 10^{11} \GeV$ were used.
Lower-red/upper-blue region is excluded by the over-production of axinos from the decay of flaon/neutralino.
Right: Broken phase.
Same $\gamma_\chi$ and $X_0$ as in the left-panel were used, and $m_\chi=500 \GeV$ were used.
Red lines are obtained by \eq{neutralino-axino-no-TI}, and correspond to $\Omega_\axino/\Omega_{\rm CDM} = 0.1, 1$ from top to bottom with $\Omega_{\rm CDM} = 0.268$.
}
\end{figure}

The axion coupling constant may be upper-bounded such as \cite{Hiramatsu:2012gg}
\beq
X_0 \lesssim 10^{11} \GeV
\eeq
Then, in addition to the the contribution of cold axion, a right amount of relic density can be achieved by axinos in a narrow parameter space as shown in Fig.~\ref{fig:axinos}.
Note that, in the symmetric phase where thermal inflation takes place, the mass of axino is contrainted to be comparable to or less than $\mathcal{O}(1) \GeV$.
On the other hand, in the broken phase, axino can be heavier, but at the price of low reheating temperature of primordial inflation.

\subsection{RH-sneutrinos}
As a scalar field, RH-sneutrinos can be produced via a coherent oscillation as described in section~\ref{sec:cosmo}, and from the decay of neutralinos similarly to the case of axinos.

\subsubsection{RH-sneutrinos from the coherent oscillation}
In the broken phase, the late-time abundance of RH-sneutrino, associated with the coherent oscillation is given by \eq{Y-RHsneu-ini}.
On the other hand, in the symmetric phase, if inflaton decays before thermal inflation begins, using \eqs{Y-RHsneu-ini}{TI-dilution}, one finds the late time abundance of $\sneu$ given by
\beq \label{Y-RHsneu-TI}
Y_{\tilde{N}} \sim \frac{\pi^2 \ g_{*S}(T_{\rm c})}{30} \frac{T_{\rm c}^3 T_{\rm d}}{V_0} \l( \frac{\tilde{N}_0}{M_\pl} \r)^2 \frac{T_{\rm R}}{m_{\tilde{N}}} 
\eeq
where $V_0^{1/4} \lesssim T_{\rm R} \lesssim \l( m_{\tilde{N}} M_\pl \r)^{1/2}$.
\begin{figure}[h]
\centering
\includegraphics[width=0.45\textwidth]{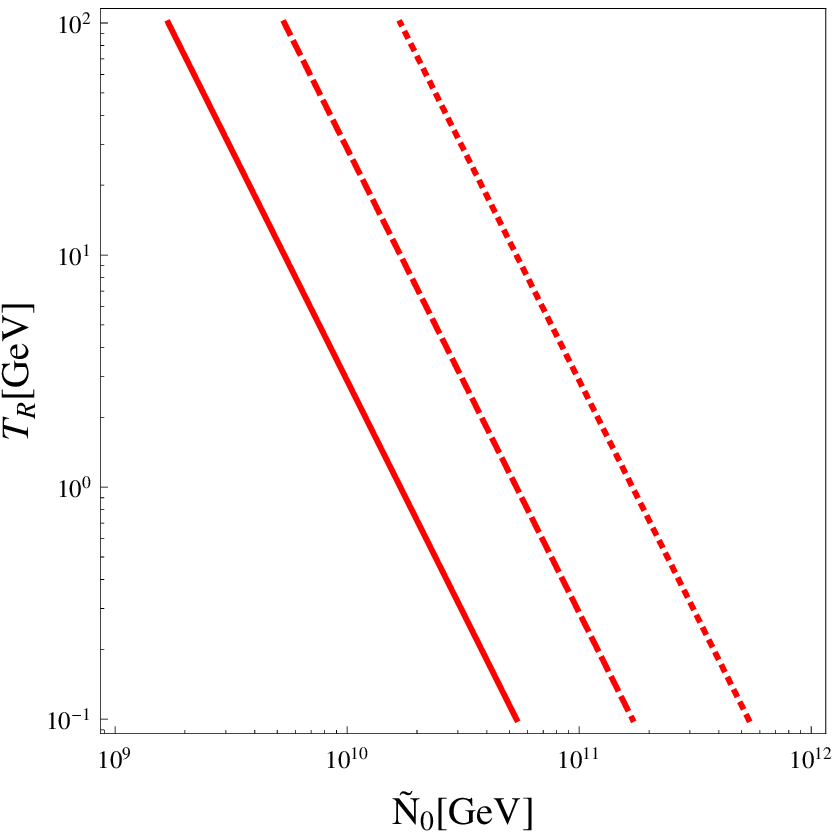}
\includegraphics[width=0.45\textwidth]{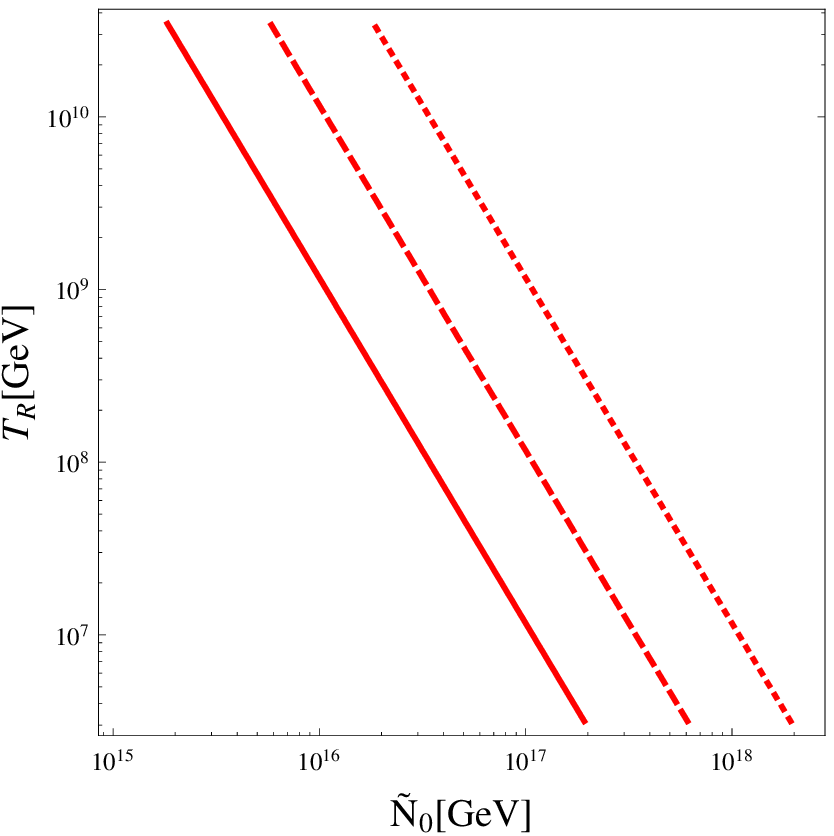}
\caption{\label{fig:RHsneu} 
Relic abundance RH-sneutrinos from a coherent oscilaltion for $m_\sneu=500 \GeV$.
Left: Broken phase.
Right: Symmetric phase.
$\mu=500 \GeV$, $m_{\rm PQ}=50 \GeV$, $X_0=2\times 10^{11} \GeV$ and $T_{\rm c} = 10 m_{\rm PQ}$ were used.
Red lines in both panels are for $Y_\sneu= 10^{-19},10^{-18},10^{-17}$ from left to right.
}
\end{figure}
Fig.~\ref{fig:RHsneu} shows $Y_\sneu$ from \eqs{Y-RHsneu-ini}{Y-RHsneu-TI} as a function of $\tilde{N}_0$ and $T_{\rm R}$.
As shown in the figure, in the broken phase intermediate scale $\sneu_0$ gives $Y_\sneu \sim \mathcal{O}(10^{-19}-10^{-17})$ which looks negligibly small, but interestingly such a small abundance of RH-sneutrino can have an observable cosmic neutrino background, as discussed in section~\ref{sec:neu-flux}.
Note that an intermediate scale $\sneu_0$ can be naturally obtained if we add $\lambda_N (N^c)^4/M_*$ with $\lambda_N \sim \mathcal{O}(10^{-2}-1)$ in the superpotential of \eq{W}.

\subsubsection{RH-sneutrinos from the neutralino decay}
In this case, one can apply the same argument as the case of axino, but with a suppression factor, 
\bea
\frac{\Gamma_{\chi \to \sneu + \nu}}{\Gamma_{\chi \to \axino + \rm SM}} 
&=& \frac{| \lambda_{\nu, \rm eff} \Theta_{\tilde{H}_u}^\chi |^2}{\gamma_\chi} \l( \frac{X_0}{m_\chi} \r)^2 \l( 1 - \frac{m_\sneu^2}{m_\chi^2} \r)^2 
\nonumber \\
&=& 5.6 \times 10^{-9} \frac{|\Theta_{\tilde{H}_u}^\chi |^2}{\gamma_\chi}
\l( \frac{\lambda_{\nu, \rm eff}}{3 \times 10^{-13}} \r)^2 
\nonumber \\
&& \times \l( \frac{X_0}{2 \times 10^{11} \GeV} \r)^2 \l( \frac{800 \GeV}{m_\chi} \r)^2 \l( 1 - \frac{m_\sneu^2}{m_\chi^2} \r)^2
\eea
where we used \eqs{chi-to-axino}{chi-to-sneutrino}.
As the result, one finds that the abundance of RH-sneutrino from neutralinos is given by
\beq
Y_\sneu \simeq \frac{\Gamma_{\chi \to \sneu + \nu}}{\Gamma_{\chi \to \axino + \rm SM}} Y_\axino \ {\rm or} \ \Omega_\sneu = \frac{m_\sneu}{m_\axino} \frac{\Gamma_{\chi \to \sneu + \nu}}{\Gamma_{\chi \to \axino + \rm SM}} \Omega_\axino
\eeq
with $\Omega_\axino$ given by \eq{neutralino-axino-no-TI}.
Note that this contribution exists irrespective of the one from coherent oscillation, and, if $Y_\axino \sim 10^{-10}$ with $m_\axino \sim 1 \GeV$, saturating the relic density, one obtains again $Y_\sneu \sim \mathcal{O}(10^{-19}-10^{-18})$ relevant for an observable neutrino flux.

A remark is in order here.
One may consider the case of RH-sneutrino LSP and axino NLSP.
In this case, if  $\sneu$ starts to oscillate with a Planckian initial oscillation amplitude in the radiation dominated universe and the reheating temperature of thermal inflation is well below $\GeV$ scale, one can have a right amount of dark matter and a small enough amount of axino NLSP so as to be consistent with observed cosmic neutrino flux.
However, the Planckian initial oscillation amplitude of $\sneu$ and sub-$\GeV$ $T_{\rm d}$ of thermal inflation is unlikely to be obtained.  
Otherwise, axino NLSP is expected to be produced too much and cause too much cosmic neutrino flux or the relic density of RH-sneutrino dark matter is not enough.
Another possibility is the case of RH-sneutrino LSP and neutralino NLSP.
In this case, one can obtain a right amount of RH-sneutrino dark matter from the coherent production by adjusting the reheating temperature of primordial inflation.
In order not to overproduce neutralino, axino may have to decay when neutralino is still in thermal bath.
This case is quite boring since it seems difficult to have any direct/indirect signature of dark matter and collider experiments may trace only NLSP if possible.

\section{Dark radiation} \label{sec:DR}
\begin{figure}[h]
\centering
\includegraphics[width=0.7\textwidth]{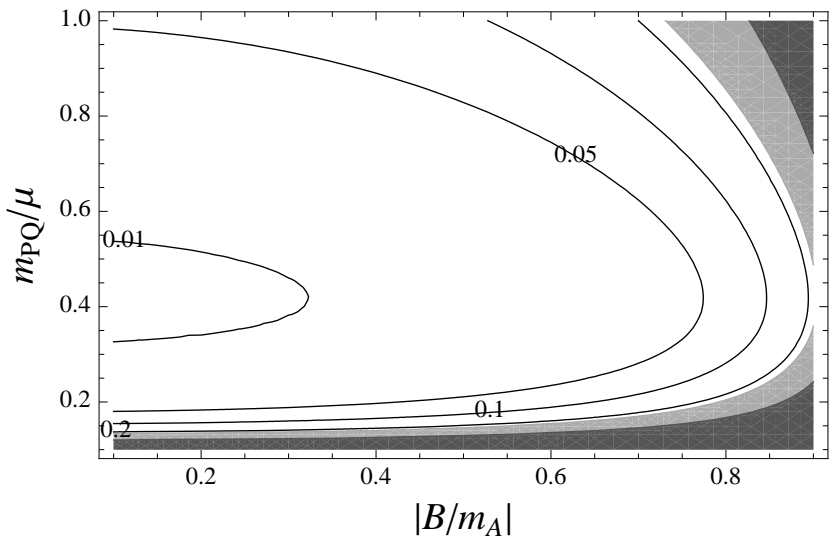}
\caption{\label{fig:DR} 
$\Delta N_{\nu, \rm eff}$ as a function of $|B/m_A|$ and $m_{\rm PQ}/\mu$ for $m_h=125 \GeV$ and $\mu=500 \GeV$.
Black lines correspond to $\Delta N_{\nu, \rm eff} = 0.01, 0.05, 0.1, 0.2$.
Gray and dark-gray regions correspond to $N_{\nu, \rm eff} \geq 3.28, 3.28 + 0.28$, respectively.
}
\end{figure}

Recent analysis based on data from Planck satellite mission showed that the relativistic degrees of freedom around the epoch of CMB decoupling is $N_{\nu, \rm eff} = 3.28 \pm 0.28$ \cite{Cooke:2013cba}. 
This may imply the existence of dark radiation, the extra relativistic degree of freedom other then SM photons and neutrinos. 
In our scenario, the axion kinetic term allows for flaton field $X$ to decay to relativistic axions contributing to the dark radiation.
In particular, the contribution can be sizable in case of thermal inflation.
Using \eq{XtoSM}, one finds (see also \cite{Choi:1996vz,Jeong:2012np})
\bea
\Delta N_{\nu, \rm eff} 
&=& \frac{\rho_a}{\rho_\nu} = \l( \frac{11}{4} \r)^{4/3} \l( \frac{g_*(T\simeq 1\MeV)}{2} \r) \l. \frac{\rho_a}{\rho_{\rm r}} \r|_{t_{\rm BBN}} 
\nonumber \\
&\simeq& \l( \frac{11}{4} \r)^{4/3} \l( \frac{g_*(T\simeq 1\MeV)}{2} \r) \l( \frac{g_{*S}(T\simeq 1\MeV)}{g_{*S}(T_{\rm d})} \r)^{1/3} \frac{\Gamma_{X \to aa}}{\Gamma_{X \to \rm SM}} 
\nonumber \\
&\simeq& \frac{10.8}{16} \l( 1 - \frac{|B|^2}{m_A^2} \r)^{-2} \l( \frac{m_{\rm PQ}}{\mu} \r)^4 f^{-1}(m_h^2/m_{\rm PQ}^2)
\eea
where we assumed $1 \GeV \lesssim T_{\rm d} < 4 \GeV$, and used $g_*(T\simeq 1 \MeV) = g_{*S}(T \simeq 1 \MeV) = 10.75$ and $g_{*S}(T_{\rm d}) = 75.75$.
As can be seen in Fig~\ref{fig:DR}, where $\Delta N_{\nu, \rm eff}$ is depicted as a function of $|B/m_A|$ and $m_{\rm PQ}/\mu$, we can have a sizable axion dark radiation in regions of $m_{\rm PQ}/\mu \sim 0.1$ and $|B|/m_A \sim 1$ where the branching fraction of flaton decay into axions is sizable.
%

%

\section{Neutrino flux} \label{sec:neu-flux}
In our scenario, NLSP can decay to LSP, producing an energetic neutrino (i.e., $\sneu \to \axino + \nu$).
The energy spectrum of neutrinos produced in the decay of NLSPs is 
\beq
\frac{d N_\nu}{d E} = \delta (E-E_{\rm ini})
\eeq
where $E_{\rm ini}$ is the energy of neutrino when it is produced.
Ignoring the mass of neutrino, one finds
\beq
E_{\rm ini} = \frac{m_\sneu}{2} \l( 1 - \frac{m_\axino^2}{m_\sneu^2} \r)
\eeq
For $\tau_\sneu < \tau_0$ with $\tau_\sneu$ and $\tau_0$ being respectively the life-time of RH-sneutrino and the age of our universe, the present cosmic neutrino flux from the decay of RH-sneutrino is given by 
\beq
\Phi_\nu(E) = \frac{1}{4 \pi} \frac{Y_\sneu s_0}{\tau_\sneu E} \l[ \frac{e^{-t/\tau_\sneu} D_\nu(E, z(t))}{H(t)} \r]_{1+z(t)=E_{\rm ini}/E}
\eeq
where $s_0$ is the entropy density at present, $1+z(t) \equiv a_0/a(t)$ with $a_0$ and $a(t)$ being respectively the scale factor at present and a time $t$, $D_\nu(E,z(t))$ is the damping factor caused by the scattering to background particles, and $H(t)$ is the expansion rate.
%
We can set $D_\nu(E, z(t)) = 1$ since neutrinos are expected to be produced at very late time $\tau_\sneu \sim \mathcal{O}(10^{-2}-1) \tau_0$ \cite{Ema:2013nda}. 
%
\begin{figure}[h]
\centering
\includegraphics[width=0.7\textwidth]{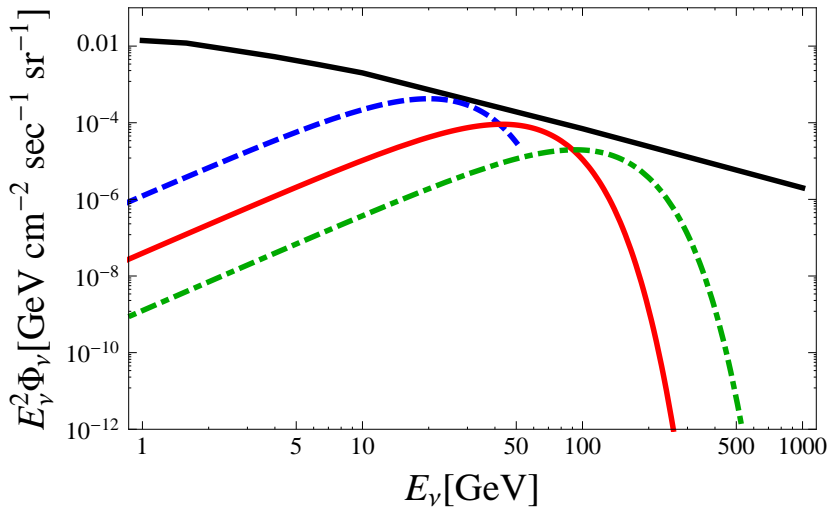}
\caption{\label{fig:neutrino-flux} 
Neutrino flux coming from the decay of NLSP (colored lines).
Thin solid black line is the conventional cosmic neutrino background \cite{Honda:2006qj}.
Dot-dashed green line is for $m_\sneu = 10 \TeV, Y_\sneu = 10^{-19}$ and $\tau_\sneu = 10^{15} {\rm sec}$. 
Solid red line is for $m_\sneu = 1 \TeV, Y_\sneu = 10^{-18}, \tau_\sneu = 10^{16} {\rm sec}$.
Dashed blue line is for $m_\sneu = 100 \GeV, Y_\sneu = 10^{-17}, \tau_\sneu = 10^{17} {\rm sec}$.
}
\end{figure}
Fig.~\ref{fig:neutrino-flux} shows neutrino flux coming from the decay of RH-sneutrino NLSPs in our scenario.
Conventional background flux of $\nu_e$ \cite{Honda:2006qj} which is consistent with the recent data \cite{Aartsen:2012uu} is also depicted in the figure.
We notice that for $m_\sneu = \mathcal{O}(0.1-10) \TeV$ a sizable neutrino flux within the reach of currently on-going or near-future experiments (such as IceCube \cite{icecube} ) can be produced if $\mathcal{O}(10^{-19}) \lesssim Y_\sneu \lesssim \mathcal{O}(10^{-17})$.
This is exactly what we expect as the would-be relic abundance of RH-sneutrino as described in section~\ref{sec:DM}.

\section{Conclusions} \label{sec:con}
In this paper, we proposed a simple supersymmetric extension of the standard model, in which both the MSSM $\mu$-term and small Yukawa couplings for the tiny Dirac neutrino mass term are simultaneously generated by the intermediate scale vacuum expectation value of PQ-field which breaks $U(1)_{\rm PQ}$ symmetry.

It was shown that a right amount of relic density can be obtained by axino LSP produced in the decay of saxion (flaton responsible for thermal inflation) and/or thermally generated neutralinos.
The possibility of right-handed sneutrino LSP which can saturate the observed relic density was also discussed.
Additionally, it was shown that in the case of thermal inflation a sizable amount of axion dark radiation, which match to the recent data from Planck satellite mission, can also be obtained in a wide range of parameter space.

Interestingly, we found that right-handed sneutrino NLSP decaying to axino LSP can produce a cosmological neutrino flux which may be observable in the near future experiments in the energy range of $\mathcal{O}(10-100) \GeV$.
This is a very unique possibility of supporting our scenario although there is still a large room of parameter space without any observable signatures.


\acknowledgments

WIP thanks Eung Jin Chun and Pyungwon Ko for valuable comments on the manuscript.
WIP is supported in part by National Research Foundation of Korea (NRF) Research Grant 2012R1A2A1A01006053.

%


\end{document}